# A Low-Cost Robust Distributed Linearly Constrained Beamformer for Wireless Acoustic Sensor Networks with Arbitrary Topology

Andreas I. Koutrouvelis, Thomas W. Sherson, Richard Heusdens and Richard C. Hendriks

*Abstract*—We propose a new robust distributed linearly constrained beamformer which utilizes a set of linear equality constraints to reduce the cross power spectral density matrix to a block-diagonal form. The proposed beamformer has a convenient objective function for use in arbitrary distributed network topologies while having identical performance to a centralized implementation. Moreover, the new optimization problem is robust to relative acoustic transfer function (RATF) estimation errors and to target activity detection (TAD) errors. Two variants of the proposed beamformer are presented and evaluated in the context of multi-microphone speech enhancement in a wireless acoustic sensor network, and are compared with other state-of-the-art distributed beamformers in terms of communication costs and robustness to RATF estimation errors and TAD errors.

## I. INTRODUCTION

BEAMFORMING (see e.g., [1]–[3] for an overview) plays an important role in multi-microphone speech enhancement [4]–[7]. The aim of a beamformer is the joint suppression of interfering noise and the preservation of an unknown target signal. The increasing usage of wireless portable devices equipped with microphones and limited power supplies, makes the notion of distributed beamforming in wireless acoustic sensor networks (WASNs) attractive compared to traditional centralized implementations [8]. The last decade, there are several proposed low-complexity distributed beamformers [9]–[18] that mainly focus on achieving a good trade-off between noise reduction and communication cost.

Both centralized and distributed beamformers typically require an estimate of the cross-power spectral density matrix (CPSDM) of the noise/noisy measurements, and estimate(s) of the relative acoustic transfer function (RATF) vector(s) of the acoustic source(s) present in the acoustic scene. Estimation errors in these quantities result in performance degradation of beamformers. Much attention has therefore been given to the development of centralized robust beamformers which minimize the effects of RATF estimation errors (see e.g., [2], [3] for an overview). Developing robust *distributed* beamformers is more challenging than developing robust centralized beamformers, as distributed beamformers cannot afford high-complexity robust solutions. Therefore, it is desired to find very low-complexity robust distributed beamformers that achieve good performance trade-offs as described previously.

A low-complexity and easily manipulated family of beamformers are those that are calculated through linearly constrained quadratic problems such as: the minimum power distortionless responce (MPDR) beamformer [19] and its multiple constrained generalization, the linearly constrained minimum power (LCMP) beamformer [20]. Both beamformers minimize the total power of the *noisy* measurements while preserving the target. Therefore, their performance highly depends on the estimation accuracy of the RATF vector of the target source [2], [3], [21]. RATF estimation errors might result in removal of the actual target source and preservation in the direction of the wrongly estimated RATF vector.

Two straightforward, low-complexity, robust alternatives to MPDR and LCMP are the minimum variance distortionless response (MVDR) beamformer [21] and the linearly constrained minimum variance (LCMV) beamformer [2], respectively. Both methods minimize the output *noise* power instead of the total noisy power, and thus require an estimate of the noise-only CPSDM. The noise CPSDM is typically estimated using a target activity detector (TAD) to identify target-free time-segments of audio. When the target is speech, this typically takes the form of a voice activity detector (see e.g., [6] for an overview). In [22], an alternative method was proposed to track the noise CPSDM also in time regions where the target is present. This method, however, highly depends on the estimation accuracy of the RATF vector of the target and its robustness to RATF estimation errors has not been tested.

Another family of low-complexity, robust alternatives to MPDR and LCMP are their diagonal loaded versions (see e.g., [23]–[25]). In both versions, the diagonal loading parameter, which is added to the main diagonal of the CPSDM, trades-off robustness against noise suppression. Specifically, by increasing the value of the diagonal loading parameter, a higher robustness to RATF estimation errors and a lower noise suppression is achieved. With diagonal loading, the use of a TAD is unnecessary. To the authors' knowledge, there are no low-complexity distributed approaches for choosing the optimal diagonal loading parameter. Additionally, a constant diagonal loading parameter will not be optimal for all acoustical scenarios and all frequency bins.

From the above it becomes clear that in addition to robustness and low-cost distributed calculations, LCMV and LCMP beamformers have the additional challenge of the RATF vector estimation of the target source and possibly the interferers. There are several centralized methods for RATF vector estimation (see e.g., [7] for an overview), however, there are yet no low-complexity distributed alternatives for arbitrary network topologies. In several applications, such as teleconferencing, the sources do not change their locations significantly over time and, therefore, one may estimate the RATF vectors of the target and/or the interferers only during initialization using a centralized approach and then use these estimated RATF vectors in the distributed beamformer. The slight positional errors that will most likely occur after this initial estimation require robust distributed beamformers. Note that in this paper, we mainly focus on this type of applications, i.e., the sources that do not significantly change their locations with respect to an initial reference location.

Notably, existing distributed beamformers can be classified based on how they address the issue of forming CPSDMs in WASNs. In the first class, the CPSDMs are approximated to form distributed implementations [9]–[12] leading to approximately optimal performance. In the second class, the proposed beamformers obtain statistical optimality but do so at the expense of restricting the topology of the underlying WASN [13]–[15]. Statistically optimal beamformers which operate in unrestricted network topologies are much less common. An early example of such a beamformer is provided in [16], based on a maximum likelihood estimated LCMP beamformer. Unfortunately, this approach suffers from scaling communication costs as the number of samples used to construct the estimated CPSDM increases. In a similar vein, in [26], a distributed beamformer based on the pseudo-coherence principle was proposed. Similar to [16], the method in [26] can operate in cyclic networks. Furthermore, the authors demonstrated how the algorithm could perform near optimally with only a finite number of iterations, resulting in low transmission complexity. More recently, in [18] a topology independent distributed beamformer (i.e. able to operate in cyclic networks) was proposed. Similar in its design to [14], this method requires very limited communication between nodes while guaranteeing convergence to the optimal beamformer. However, it was also demonstrated that the rate of this convergence was slow, requiring a large number of iterations to achieve this point. In practice, with even slowly varying sound fields such a rate of convergence may be detrimental to overall performance.

In this paper, we propose a new robust distributed linearly constrained beamformer, addressing the aforementioned challenges. The optimization problem of the proposed method nulls each interferer using a linear equality constraint, reducing the full-element noise or noisy CPSDM to a *block-diagonal* form. In contrast to MVDR, MPDR, LCMV and LCMP beamformers, the proposed objective function does not take into account correlation between different nodes in the WASN. Additionally, such an objective function is more convenient for distributed beamforming in WASNs of arbitrary topologies and significantly reduces the communication costs therein.

We show under realistic conditions, i.e., when the algorithms use non-ideally estimated RATF vectors and a non-ideal TAD, that the proposed method achieves a better predicted intelligibility than the MVDR and LCMV. The proposed method is less sensitive to RATF estimation errors, when TAD errors are not negligible, because of the block-diagonal form of the CPSDM.

The remainder of the paper is organized as follows. Section II presents the signal model. Section III reviews several methods of estimating the RATF vectors of the sources and the noisy/noise CPSDMs. Section IV reviews the centralized and distributed linearly constrained beamformers. Section V presents the centralized and distributed versions of the proposed method. Section VI shows the experimental results. Finally, concluding remarks are drawn in Section VII.

## II. Signal Model

Consider an arbitrary undirected WASN of $N$ nodes. Without loss of generality, we assume that the underlying network (which is potentially cyclic) is connected. Denote by $V = \{1, \cdots, N\}$ the set of node indices and by $E$ the set of edges of the network whereby $(i,j) \in E \iff i,j \in V, i \neq j$ can communicate with one another. Each node $\kappa$ is equipped with $M_\kappa$ microphones, where $\sum_{\kappa \in V} M_\kappa = M$, thus forming an $M$-element microphone array. One of the $M$ microphones is selected as the reference microphone for the beamforming purpose. The distributed beamformers presented in this paper are formulated in the short-time Fourier transform (STFT) domain on a frame-by-frame basis. The noisy DFT coefficient of the $j$-th ($j = 1, \cdots, M$) microphone of the $k$-th frequency bin of the $\beta$-th frame is given by

$$y_j(k,\beta) = \underbrace{a_j(k,\beta)s(k,\beta)}_{x_j(k,\beta)} + \sum_{i=1}^{r} \underbrace{b_{ij}(k,\beta)v_i(k,\beta)}_{n_{ij}(k,\beta)} + u_j(k,\beta) \quad (1)$$

with $s(k,\beta)$ and $v_i(k,\beta)$ the target source and the $i$-th interferer at the reference microphone, $a_j(k,\beta)$ and $b_{ij}(k,\beta)$ the RATF vectors elements of each with respect to the $j$-th microphone, and $x_j(k,\beta)$, $n_{ij}(k,\beta)$ and $u_j(k,\beta)$ the target source, the $i$-th interferer and ambient noise at the $j$-th microphone. Note that the reference microphone element of the RATF vectors is always equal to 1. Moreover, in the case of reverberant environments, the RATF vectors may also include a component due to early reverberation [27], [28]. Late reverberation and microphone self-noise are typically included in the ambient noise component. Note that even the late reverberation of the target has to be assigned to the ambient noise component because it reduces intelligibility [29], [30]. Thus, it should be reduced via the use of the beamformer. However, the early reflections (typically the first 50 ms [30]) are desired to be maintained because they typically contribute to intelligibility [29], [30]. Therefore, the ambient noise component is given by

$$u_j(k,\beta) = l_j^s(k,\beta) + \sum_{i=1}^{r} l_j^{v_i}(k,\beta) + c_j(k,\beta),$$

where $l_j^s(k,\beta)$ is the late reverberation component due to the target, $l_j^{v_i}(k,\beta)$ is the late reverberation component due to the $i$-th interferer, and $c_j(k,\beta)$ is the microphone self-noise.

In the sequel, we neglect the frame and frequency indices for the sake of brevity. Stacking all variables into vectors, Eq. (1) can be rewritten as

$$\mathbf{y} = \mathbf{x} + \underbrace{\sum_{i=1}^{r} \mathbf{n}_i + \mathbf{u}}_{\mathbf{n}} \in \mathbb{C}^{M \times 1}.$$

The CPSDM of $\mathbf{y}$ is given by $\mathbf{P_y} = \mathrm{E}[\mathbf{yy}^H]$, where $\mathrm{E}[\cdot]$ denotes statistical expectation. Assuming all sources are mutually uncorrelated, we have

$$\mathbf{P_y} = \mathbf{P_x} + \underbrace{\sum_{i=1}^{r} \mathbf{P}_{\mathbf{n}i} + \mathbf{P_u}}_{\mathbf{P_n}} \in \mathbb{C}^{M \times M}, \quad (2)$$

where $\mathbf{P_x} = \mathrm{E}[\mathbf{xx}^H] = p_s \mathbf{aa}^H$ and $\mathbf{P}_{\mathbf{n}i} = \mathrm{E}[\mathbf{n}_i \mathbf{n}_i^H] = p_{v_i} \mathbf{b}_i \mathbf{b}_i^H$ are the CPSDMs of the target source and the $i$-th interferer at the microphones, respectively. Note that $p_s$ and $p_{v_i}$ are the power spectral densities of the target and the $i$-th interferer, respectively. Finally, the CPSDM of the ambient noise component, $\mathbf{P_u}$, is given by

$$\mathbf{P_u} = \mathrm{E}[\mathbf{uu}^H] = \underbrace{\mathbf{P}_{\mathbf{l}s} + \sum_{i=1}^{r} \mathbf{P}_{\mathbf{l}u_i}}_{\mathbf{P_l}} + \mathbf{P_c} \in \mathbb{C}^{M \times M},$$

where $\mathbf{P_l}$ denotes the CPSDM of the late reverberation, and $\mathbf{P_c}$ the CPSDM of the microphone self-noise.

## III. ESTIMATION OF SIGNAL MODEL PARAMETERS

The CPSDMs and the RATF vectors of the sources are unknown and have to be estimated in order to be available to the beamformers discussed in the sequel. In Sections III-A and III-B, we review some existing methods for RATF vector and CPSDM estimation, respectively.

### A. Estimation of RATF Vectors

In practical applications, the true RATF vectors are reverberant due to room acoustics [28], [31], [32]. Several centralized methods have been proposed to estimate these RATF vectors (see e.g., [7] for an overview). In [28], the RATF vector of the target source is estimated by exploiting the assumption that the noise field is stationary. However, when the interferers are non-stationary, this can result in significant degradation in performance [31]. In [32] the subspaces of the target and interferers are estimated using a generalized eigenvalue decomposition (GEVD) combined with a TAD. While distributed methods have been proposed in the literature for performing GEVD-based subspace estimation in restricted network topologies (i.e., fully connected) [33], to our best knowledge, there are currently no distributed versions of the GEVD that operate in general cyclic networks.

In this work, we assume that estimates of the RATF vectors, $\hat{\mathbf{a}}$ and $\hat{\mathbf{b}}_i$, for $i = 1, \cdots, r$, are available at the initialization phase. In situations where the sources do not change their locations significantly with respect to an initial position, such as teleconferencing, the RATF vectors can be estimated (e.g., in a centralized way) during such an initialization. This will result in RATF estimation errors if the sources make some slight movements and, therefore, robust beamformers are required.

### B. Estimation of CPSDMs

The LCMP and the MPDR beamformers depend on an estimate of the noisy CPSDM, $\hat{\mathbf{P}}_{\mathbf{y}}$. Typically, this estimate is computed using the sample average, which is given by

$$\hat{\mathbf{P}}_{\mathbf{y}} = \frac{1}{|L_y|} \sum_{l_y \in L_y} \mathbf{y}(l_y) \mathbf{y}^H(l_y),$$

where $L_y$ is the set of frames of the entire time horizon and $|\cdot|$ denotes the cardinality of a set. The LCMV and the MVDR beamformers depend on an estimate of the noise CPSDM, $\hat{\mathbf{P}}_{\mathbf{n}}$. The noise CPSDM is estimated using the set of noise-only frames denoted by $L_n$, i.e.,

$$\hat{\mathbf{P}}_{\mathbf{n}} = \frac{1}{|L_n|} \sum_{l_n \in L_n} \mathbf{y}(l_n) \mathbf{y}^H(l_n),$$

where $|L_n| < |L_y|$. In order to obtain $\hat{\mathbf{P}}_{\mathbf{n}}$, a TAD is required to detect target presence/absence for each frame. The above two averages are updated in an online fashion, i.e., the average is updated for every frame using the average of the previous frame. This procedure becomes computationally demanding in a distributed context for two reasons. Firstly, the entire observation vector must be available at each time frame resulting in the need for data flooding. Secondly, that the storage of the entire CPSDM scales with the network size.

Estimation of the ambient noise CPSDM $\mathbf{P_u}$ is a difficult task due to the late reverberation CPSDM $\mathbf{P_l}$. Using a TAD it is nearly impossible to estimate $\mathbf{P_l}$ alone. For sufficiently large rooms, the late reverberation is typically modelled as an ideal spherical isotropic noise field [7], [34]. That is,

$$\hat{\mathbf{P}}_{\mathbf{l}} = \hat{p}_{\text{iso}} \mathbf{P}_{\text{iso}}, \quad (3)$$

where for the $k$-th frequency bin, the $(i,j)$-th element of $\mathbf{P}_{\text{iso}}$ is given by

$$\mathbf{P}_{\text{iso},i,j} = \mathrm{sinc}\left(\frac{2\pi k f_s d_{i,j}}{\Phi c}\right), \quad (4)$$

where $d_{i,j}$ is the distance between microphones $i$ and $j$, $f_s$ is the sampling frequency, $\Phi$ is the number of frequency bins, and $c$ is the speed of sound. The scaling $\hat{p}_{\text{iso}}$ can be estimated using several centralized methods (see e.g., [34]). To the best of our knowledge, there are no distributed methods for obtaining $\hat{p}_{\text{iso}}$.

Fig. 1 shows the values of the correlation function of Eq. (4) for various frequencies and distances $d_{i,j}$. The correlation can be roughly divided into two interesting frequency regions: one highly correlated on the left and one much less correlated on the right. The boundary between these regions occurs at the first zero-crossing given by $f_c = c/(2d_{i,j})$. It is clear that, the larger $d_{i,j}$ becomes, the smaller $f_c$ is.

The CPSDM of the microphone self-noise, $\mathbf{P_c} = c\mathbf{I}$ (where $c$ is the power at each microphone), can be estimated in silent frames only (i.e., neither target nor interferers are active).

## IV. LINEARLY CONSTRAINED BEAMFORMING

Most linearly constrained beamformers are obtained from the following general optimization problem [1], [2], [20]

$$\hat{\mathbf{w}} = \arg\min_{\mathbf{w}} \mathbf{w}^H \mathbf{P} \mathbf{w} \text{ s.t. } \mathbf{w}^H \mathbf{\Lambda} = \mathbf{f}^H, \quad (5)$$

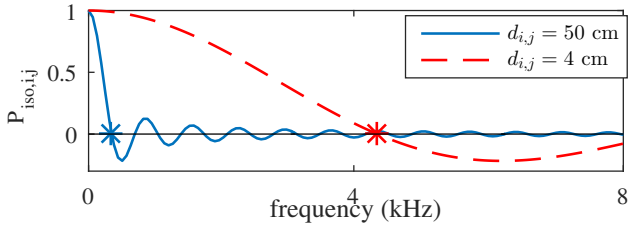

Fig. 1: The spherically isotropic noise field correlation between two microphones $i, j$ of distances $d_{i,j} = 4, 50$ cm and $f_s = 16$ kHz. The star marker denotes the first zero-crossing $f_c$.

where $\mathbf{\Lambda} \in \mathbb{C}^{M \times d}$, $\mathbf{f} \in \mathbb{C}^{d \times 1}$, and $\mathbf{P} \in \mathbb{C}^{M \times M}$ is typically the CPSDM of the noise or noisy measurements. The $d$ constraints used in the optimization problem of Eq. (5) include at least the distortionless constraint for the target source, i.e., $\mathbf{w}^H \mathbf{a} = 1$, and, commonly, the nulling of the interferers, $\mathbf{w}^H \mathbf{b}_i = 0$ [1], [32], [35]. If we assume that $r < M - 1$, the linearly constrained beamformer can null all interferers and still have control on the minimization of the objective function. In this case, $\mathbf{\Lambda}$ and $\mathbf{f}$ are given by

$$\mathbf{\Lambda} = \begin{bmatrix} \mathbf{a} & \mathbf{b}_1 & \cdots & \mathbf{b}_r \end{bmatrix}, \text{ and } \mathbf{f} = \begin{bmatrix} 1 & 0 & \cdots & 0 \end{bmatrix}^H. \quad (6)$$

It should be noted that by increasing the number of nulling constraints, the ambient output noise power may be boosted. The boost depends on the locations of the interferers [2] and the number of available degrees of freedom $(M - r - 1)$. However, in applications when $r \ll M - 1$ this impact is much less significant. If $r < M - 1$ and $\mathbf{P}$ is invertible, the optimization problem in Eq. (5), using the constraints in Eq. (6), has a closed-form solution given by [2]

$$\hat{\mathbf{w}} = \mathbf{P}^{-1} \mathbf{\Lambda} \left( \mathbf{\Lambda}^H \mathbf{P}^{-1} \mathbf{\Lambda} \right)^{-1} \mathbf{f}.$$

When $\mathbf{P} = \mathbf{P}_\mathbf{y}$, the linearly constrained beamformer takes the form of the LCMP beamformer given by

$$\hat{\mathbf{w}} = \arg\min_{\mathbf{w}} \mathbf{w}^H \mathbf{P}_\mathbf{y} \mathbf{w} \text{ s.t. } \mathbf{w}^H \mathbf{\Lambda} = \mathbf{f}^H, \quad (7)$$

while if $\mathbf{P} = \mathbf{P}_\mathbf{n}$, the LCMV is obtained and is given by

$$\hat{\mathbf{w}} = \arg\min_{\mathbf{w}} \mathbf{w}^H \mathbf{P}_\mathbf{n} \mathbf{w} \text{ s.t. } \mathbf{w}^H \mathbf{\Lambda} = \mathbf{f}^H.$$

In the sequel, when we use the acronyms LCMV and LCMP we mean the LCMV and LCMP versions with the constraints given in Eq. (6). Another interesting linearly constrained beamformer is the one that has only the ambient noise component in the objective function [36], i.e.,

$$\hat{\mathbf{w}} = \arg\min_{\mathbf{w}} \mathbf{w}^H \mathbf{P}_\mathbf{u} \mathbf{w} \text{ s.t. } \mathbf{w}^H \mathbf{\Lambda} = \mathbf{f}^H. \quad (8)$$

In this paper, we will refer to the linearly constrained beamformer in Eq. (8) as the ambient LCMV (ALCMV).

Using Eq. (2), the objective function of the LCMP problem, as noted in Eq. (7), is given by

$$\mathbf{w}^H \mathbf{P}_\mathbf{y} \mathbf{w} = p_s \mathbf{w}^H \mathbf{a} \mathbf{a}^H \mathbf{w} + \sum_{i=1}^{r} p_{v_i} \mathbf{w}^H \mathbf{b}_i \mathbf{b}_i^H \mathbf{w} + \mathbf{w}^H \mathbf{P}_\mathbf{u} \mathbf{w}.$$

Due to the included constraints in the LCMP (see Eq. (6)), the contributions of the early components of the sources to the objective function of Eq. (7) are constant. Thus, if $\hat{\mathbf{P}}_\mathbf{y} = \mathbf{P}_\mathbf{y}$, $\hat{\mathbf{P}}_\mathbf{n} = \mathbf{P}_\mathbf{n}$, $\hat{\mathbf{P}}_\mathbf{u} = \mathbf{P}_\mathbf{u}$, and $\hat{\mathbf{\Lambda}} = \mathbf{\Lambda}$, the LCMP, LCMV and ALCMV beamformers are all equivalent. In practice, this never happens as there are always RATF estimation errors and CPSDM estimation errors, as explained previously.

### A. RATF estimation errors

There are two interesting cases. In the first case, if $\hat{\mathbf{P}}_\mathbf{y} = \mathbf{P}_\mathbf{y}$, $\hat{\mathbf{P}}_\mathbf{n} = \mathbf{P}_\mathbf{n}$, and $\hat{\mathbf{a}} = \mathbf{a}$, LCMP is equivalent to LCMV [2]. However, if $\hat{\mathbf{a}} \neq \mathbf{a}$, the LCMV beamformer (provided that $\hat{\mathbf{P}}_\mathbf{n}$ is accurately estimated), is more robust than the LCMP [2]. This is because LCMP will try to remove the actual target related to the RATF $\mathbf{a}$ as this is included in $\mathbf{P}_\mathbf{y}$, while the preservation constraint is on the wrongly estimated $\hat{\mathbf{a}}$. However, if there are also TAD errors, $\hat{\mathbf{P}}_\mathbf{n}$ may also contain portions of $\mathbf{P}_\mathbf{x}$ and, as a result, the LCMV may also have severe performance degradation like the LCMP.

In the second case, if $\hat{\mathbf{P}}_\mathbf{n} = \mathbf{P}_\mathbf{n}$, $\hat{\mathbf{P}}_\mathbf{u} = \mathbf{P}_\mathbf{u}$, and $\hat{\mathbf{b}}_i = \mathbf{b}_i$, for $i = 1, \cdots, r$, LCMV is equivalent to ALCMV. However, if any of the $\hat{\mathbf{b}}_i$'s contain estimation errors, there will be power leakage of the corresponding interferer(s), which is not controllable, neither by the objective function nor by the constraints of the ALCMV problem in Eq. (8). Moreover, if there are interferers whose RATF vectors have not been placed in the constraints, the ALCMV will also be unable to reduce them in a controlled way. In contrast, if $\hat{\mathbf{P}}_\mathbf{n}$ is estimated accurately, the LCMV will reduce these power leakages. In this case, the LCMV will most likely have a better noise reduction performance than its ALCMV counterpart.

We can conclude that the performance degradation of linearly constrained beamformers due to RATF estimation errors is mainly influenced by the selection of the CPSDM, $\mathbf{P}$, in the objective function of Eq. (5). A low-cost robust linearly constrained beamformer should have good performance under both RATF estimation errors and TAD errors. There are several approaches to achieve this. The most popular is via diagonal loading of $\mathbf{P}$. However, to the authors' knowledge there are no low-cost distributed approaches for optimally selecting the diagonal loading value. Another robust low-cost option is to use a fixed superdirective linearly constrained beamformer, i.e., a linearly constrained beamformer with a (semi)fixed $\mathbf{P}$ [5]. A fixed linearly constrained beamformer does not use a TAD and guarantees that there will not be any portion of $\mathbf{P}_\mathbf{x}$ in $\mathbf{P}$. Two interesting fixed linearly constrained beamformers are discussed in the next section.

### B. Fixed Superdirective Linearly Constrained Beamformers

The fixed superdirective beamformers [5] assume a certain noise field and use in the objective function a certain coherence function like the one in Eq. (3). Since the early components of the interferers can be nullified using a linearly constrained beamformer, the noise field that remains is the late reverberation as explained previously in this section. Recall from Section III-B, that the estimation of $\mathbf{P}_\mathbf{u}$ is a difficult task due to the CPSDM of the late reverberation, $\mathbf{P}_\mathbf{l}$. Typically, in the literature (see e.g., [5], [37], [38]) models of $\mathbf{P_l}$ are used in beamformers instead. The most common choice is to use $\mathbf{P_{iso}}$. If one chooses $\mathbf{P} = \mathbf{P_{iso}}$, the microphone self-noise will be boosted in low frequencies [5]. Thus, a diagonal-loaded version is typically used [5], [39], i.e.,

$$\hat{\mathbf{w}} = \arg\min_{\mathbf{w}} \mathbf{w}^H (p_{\text{iso}}\mathbf{P}_{\text{iso}} + \mathbf{P_c})\mathbf{w} \text{ s.t. } \mathbf{w}^H \mathbf{\Lambda} = \mathbf{f}^H, \quad (9)$$

where $\mathbf{P_c} = c\mathbf{I}$ (see Section III-B). Although, the microphone-self noise power, $c$, typically remains constant over time, $p_{\text{iso}}$ changes. To the best of our knowledge, there are no distributed estimation methods of the scaling coefficient $p_{\text{iso}}$. We call the beamformer in Eq. (9) as isotropic LCMV (ILCMV).

Another popular fixed linearly constrained beamformer uses in the objective function the most simplistic option which is $\mathbf{P} = \mathbf{I}$, i.e.,

$$\hat{\mathbf{w}} = \arg\min_{\mathbf{w}} \mathbf{w}^H \mathbf{w} \text{ s.t. } \mathbf{w}^H \mathbf{\Lambda} = \mathbf{f}^H. \quad (10)$$

In this paper, we will refer to this as the linearly constrained delay and sum (LCDS) beamformer. It is identical to the fixed beamformer of the generalized side-lobe canceller implementation of the LCMP beamformer (using the constraints in Eq. (6)) in [32]. Unlike ILCMV, the LCDS is easily distributable due to the separable nature of the objective function. This can be achieved via similar methods to those demonstrated in Section V-C and need only be performed once. Following this, the output can be computed via data aggregation or by solving a simple averaging problem, again lending itself to distributed implementations.

Similar to ALCMV, the ILCMV and LCDS beamformers cannot control power leakages due to inaccurate estimates of the interferers' RATF vectors and cannot control interferers which are not included in the constraints.

### C. Other Related Linearly Constrained Beamformers

If we skip the nulling constraints and only impose the target distortionless constraint, the LCMV (LCMP) reduces to the MVDR (MPDR) [1], [19]. Similar to LCMV and LCMP, MVDR and MPDR are equivalent under the assumption that $\hat{\mathbf{P}}_\mathbf{y} = \mathbf{P_y}$ and $\hat{\mathbf{P}}_\mathbf{n} = \mathbf{P_n}$ and $\hat{\mathbf{a}} = \mathbf{a}$ [2]. However, when $\hat{\mathbf{a}} \neq \mathbf{a}$, the MVDR is more robust to RATF estimation errors [2], [21]. A special case of the MPDR is the delay and sum (DS) beamformer [27] which replaces the noisy CPSDM with the identity matrix. The DS has worse performance compared to the MVDR (MPDR) in correlated noise fields but results in very robust performance to RATF estimation errors [21] and TAD errors.

### D. Distributed Linearly Constrained Beamformers

The development of distributed beamformers has focused on adapting LCMV (LCMP) based approaches for use in WASNs. However, this adaptation has not come without additional challenges [40]. Most notable is the limited communication between devices which makes the formation of estimated CPS-DMs nearly impossible without the use of a fusion center [8]. To address this, two main classes of distributed beamformers have appeared in the literature: approximately optimal variants and optimal approaches which operate in certain networks.

One such sub-optimal variant is the distributed DS beamformer introduced in [9]. Based on randomised gossip [41], this low-cost method operates in general cyclic networks but fails to exploit spatial correlation to improve noise reduction. In contrast, distributed approximations of the MVDR beamformer [10], [11] assume that disjoint nodes are uncorrelated essentially masking the true CPSDMs. While lending themselves to distributed implementations, such approaches fail to take into account the true correlations between observed signals across the network, resulting in sub-optimal performance.

By restricting the network topology, typically to be acyclic or fully connected, optimal distributed beamformers have been proposed. These algorithms [14], [15] exploit efficient data aggregation to construct global beamformers from a composition of local filters and have been shown to be iteratively optimal. However, the additional communication overhead required to maintain a constant network topology across frames can be prohibitively expensive due to unpredictable network dynamics. Furthermore, such maintenance may be impossible in the case of node failure.

It is worth mentioning that it is not the use of an acyclic network in [14], [15] itself which is limiting, but rather the need for this network to be invariant over time. In [18], this point was exploited to form a fully distributed beamformer for use in general cyclic topologies. Like [14] and [15], [18] constructs a global beamformer as a composition of local beamformers at each node. Importantly, the method by which these local beamformers are combined does not depend on the underlying network topology. This allows the network to vary between frames, overcoming the need for maintaining a fixed topology in all time instances. The method in [18] was shown to be iteratively optimal with its main drawback being a decrease in convergence rate compared to [14], requiring a larger number of frames to obtain near optimal performance.

In contrast, in [16], an optimal distributed beamformer was proposed for use in cyclic networks by exploiting the structure of estimated CPSDMs to cast LCMP beamforming as distributed consensus. However, for CPSDM estimates based on a large number of frames, the proposed algorithm's communication cost scaled poorly. In contrast to [13]–[15] and [18], a benefit of [16] was that the proposed implementation was frame-optimal, i.e. that it obtained the performance of an equivalent centralized implementation in each frame. The beamformer proposed in [26] exploited a similar method of distributed implementation, but exploited the pseudo coherence principle of human speech to overcome the scaling communication costs found in [16].

The approaches of both [16] and [26] made use of internal optimization schemes which require a large number of iterations per frame to obtain optimal performance. However, in [26] it was shown that near optimal performance could be obtained using only a finite number of iterations of this internal solver. Such a result raises the question whether a similar approach could be employed as a general way of reducing the transmission costs associated with cyclic beamforming methods. For the beamformers proposed in this work, this point is touched upon in Section V-G.

In contrast to the methods above, the beamformers proposed in Section V are fully distributable without imposing restrictions on the underlying network topology or scaling communication costs while also being optimally computable in each frame. In this way, the proposed methods combine the strengths of existing distributed beamformers while also avoiding their various limitations.

## V. PROPOSED METHOD

In the previous section, we have highlighted the susceptibility of several existing beamformers to RATF estimation errors and TAD errors and the challenge of deploying these algorithms in distributed contexts. Here, we propose two different linearly constrained beamformers which are efficiently distributable for arbitrary network topologies, robust to RATF estimation errors and TAD errors, while at the same time are able to control the power leakage of the interferers.

Typically, the microphones within a node are nearby, while the microphones from different nodes are further away. Therefore, the late reverberation will be highly correlated in the first case, while in the latter less correlated (see Fig. 1). Therefore, providing that the nodes are sufficiently far away from each other, one may approximate the full element matrix $\mathbf{P_u}$ with the *block-diagonal* matrix $\bar{\mathbf{P}}_\mathbf{u}$ where every block corresponds to the CPSDM of the late reverberation of one node only and the microphone-self noise. Therefore, we propose the block-diagonal ALCMV (BDALCMV) which is given by

$$\hat{\mathbf{w}} = \arg\min_{\mathbf{w}} \mathbf{w}^H \bar{\mathbf{P}}_\mathbf{u} \mathbf{w} \text{ s.t. } \mathbf{w}^H \mathbf{\Lambda} = \mathbf{f}^H. \quad (11)$$

Note that if every node has only one microphone, $\bar{\mathbf{P}}_\mathbf{u}$ becomes diagonal. This block-diagonalization lends itself to distributed implementations, reflecting a similar objective structure to that of the DS and LCDS beamformer.

While the proposed BDALCMV beamformer has a number of benefits from the perspective of distributed signal processing, like ALCMV, the challenge becomes the estimation of $\bar{\mathbf{P}}_\mathbf{u}$, and handling the possible power leakages of the interferers as in the case of DS, LCDS, ALCMV. Therefore, in Sections V-A, and V-B we introduce two variations of the BDALCMV beamformer which do not require the estimation of $\bar{\mathbf{P}}_\mathbf{u}$ and are robust to power leakages of the interferers. Moreover, in Sections V-C—V-G, we introduce distributed implementations of the proposed beamformers.

### A. BDLCMP Beamformer

The first proposed practical variant of BDALCMV is the BDLCMP which uses in the objective function the block-diagonal noisy CPSDM, $\bar{\mathbf{P}}_\mathbf{y}$. That is,

$$\hat{\mathbf{w}} = \arg\min_{\mathbf{w}} \mathbf{w}^H \bar{\mathbf{P}}_\mathbf{y} \mathbf{w} \text{ s.t. } \mathbf{w}^H \mathbf{\Lambda} = \mathbf{f}^H. \quad (12)$$

This results in a local estimation problem, which can be carried out independently at each node without the need of a TAD. This method handles the possible power leakages due to inaccurate estimates of the interferers' RATF vectors and can suppress the interferers that are not included in the constraints.

In case of RATF estimation errors of the target source, the BDLCMP will have similar problems to the LCMP because in the block-diagonal matrices, there will be portions of the corresponding target block-diagonal CPSDMs. However, the performance degradation will not be that great as with the LCMP. This can be easily explained by considering the extreme scenario of a fully correlated noise field in which we assume that $M > r+1$, $\hat{\mathbf{P}}_\mathbf{y} = \mathbf{P_y}$, $\mathbf{P_u} \approx 0$, $\hat{\mathbf{b}}_i = \mathbf{b}_i, i = 1, \cdots, r$ and $\hat{\mathbf{a}} \neq \mathbf{a}$. In this case, the optimization problem of LCMP in Eq. (7) will be approximately equivalent[1] to the following optimization problem:

$$\hat{\mathbf{w}} = \arg\min_{\mathbf{w}} \mathbf{w}^H \hat{\mathbf{P}}_\mathbf{y} \mathbf{w} \text{ s.t. } \mathbf{w}^H \tilde{\mathbf{\Lambda}} = \tilde{\mathbf{f}}^H,$$

where

$$\tilde{\mathbf{\Lambda}} = \begin{bmatrix} \hat{\mathbf{a}} & \mathbf{a} & \hat{\mathbf{b}}_1 & \cdots & \hat{\mathbf{b}}_r \end{bmatrix}, \text{ and } \tilde{\mathbf{f}}^H = \begin{bmatrix} 1 & 0 & 0 & \cdots & 0 \end{bmatrix}.$$

That is, the LCMP will approximately nullify the target source. In contrast, due to the block-diagonal CPSDM, the BDLCMP will approximately nullify the target source iff $M > rN + 2r + 1$, where $N$ is the number of nodes. Specifically, if $M > rN + 2r + 1$ is satisfied, the BDLCMP will be approximately equivalent to the following optimization problem:

$$\hat{\mathbf{w}} = \arg\min_{\mathbf{w}} \mathbf{w}^H \hat{\bar{\mathbf{P}}}_\mathbf{y} \mathbf{w} \text{ s.t. } \mathbf{w}^H \tilde{\mathbf{\Lambda}} = \tilde{\mathbf{f}}^H,$$

where

$$\tilde{\mathbf{\Lambda}} = \begin{bmatrix} \hat{\mathbf{a}} & \tilde{\mathbf{a}}_1 & \tilde{\mathbf{a}}_2 & \cdots & \tilde{\mathbf{a}}_N & \hat{\mathbf{b}}_1 \cdots \hat{\mathbf{b}}_r & \tilde{\mathbf{b}}_{11} \cdots \tilde{\mathbf{b}}_{1N} \cdots \tilde{\mathbf{b}}_{r1} \cdots \tilde{\mathbf{b}}_{rN} \end{bmatrix},$$
$$\tilde{\mathbf{f}}^H = \begin{bmatrix} 1 & 0 & 0 & \cdots & 0 \end{bmatrix}$$
$$\tilde{\mathbf{a}}_i = \begin{bmatrix} \mathbf{0} & \mathbf{a}_i & \mathbf{0} \end{bmatrix}^H, \quad \tilde{\mathbf{b}}_{ji} = \begin{bmatrix} \mathbf{0} & \mathbf{b}_{ji} & \mathbf{0} \end{bmatrix}^H \in \mathbb{C}^{M \times 1}.$$

Here $\mathbf{a}_i, \mathbf{b}_{ji}$ are the elements of the RATF vector $\mathbf{a}, \mathbf{b}_j$ corresponding to node $i$, respectively. Note that for $M < rN + 2r + 1$ the BDLCMP will not have enough degrees of freedom to achieve $\mathbf{w}^H \tilde{\mathbf{a}}_i = 0$ ($i = 1, \cdots, N$) and, thus, will not nullify the target signal. Thus, more microphones are needed in the BDLCMP beamformer to nullify the target signal compared to the LCMP beamformer. Hence, the BDLCMP is more robust to target RATF estimation errors compared to the LCMP for the same number of microphones $M$, when $M < rN + 2r + 1$, in this particular scenario of a fully correlated noise field. In more general noise fields, where $\mathbf{P_u}$ is not negligible, both LCMP and BDLCMP will not nullify the target using the same finite number of microphones. However, LCMP will suppress more the target signal than the BDLCMP, because the first exploits the full-element noisy CPSDM matrix.

Fig. 2 shows the directivity patterns of LCMP and BDLCMP for a simple acoustic scenario with a linear microphone array separated into two nodes where each node has three microphones. The target source is at $80°$, but the estimated RATF vector of the target is at $90°$. The interferers and their RATF vectors are at $10°, 50°$ and $160°$. All RATF vectors are anechoic in this example and there is a slight amount of microphone-self noise. It is clear from the directivity pattern in Fig. 2, that LCMP suppresses the target signal significantly, while BDLCMP does not.

---

[1]It is approximately equivalent because $\mathbf{P_u} \approx 0$. Moreover, the target RATF estimation errors should be sufficiently large.

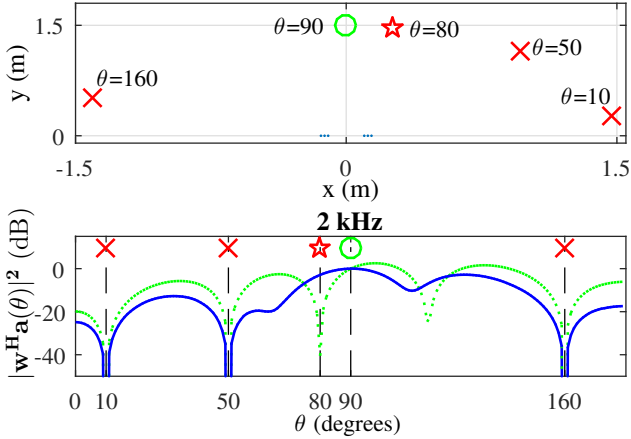

Fig. 2: Example: three interferers (with marker 'x') and one target (with marker ⋆) at $80°$. The RATF vector of the target points at $90°$. The directivity pattern, $|\mathbf{w}^H\mathbf{a}(\theta)|^2$ (in dB), is computed in the range $0° \leq \theta \leq 180°$, for BDLCMP (solid line) and LCMP (dotted line), for the frequency 2 kHz.

It is worth mentioning that if $\hat{\mathbf{b}}_i \neq \mathbf{b}_i$, it easy to show (following the same steps as before) that the LCMP will typically suppress more the $i$-th interferer than BDLCMP, if both use the same number of microphones. This means that the power leakages of the interferers will be suppressed more with the LCMP compared to the BDLCMP. Nevertheless, we will experimentally show in Section VI, that the final intelligibility improvement of BDLCMP is much greater than the LCMP, because BDLCMP distorts much less the target.

### B. BDLCMV Beamformer

To further increase the robustness of the proposed method, we introduce the BDLCMV variant which uses in the objective function the block-diagonal version of the noise CPSDM, $\bar{\mathbf{P}}_\mathbf{n}$. Therefore, the BDLCMV is given by

$$\hat{\mathbf{w}} = \arg\min_{\mathbf{w}} \mathbf{w}^H \bar{\mathbf{P}}_\mathbf{n} \mathbf{w} \text{ s.t. } \mathbf{w}^H \mathbf{\Lambda} = \mathbf{f}^H. \quad (13)$$

Similar to the relationship between LCMV and LCMP, the BDLCMV typically enjoys more robustness than the BDLCMP when $\bar{\mathbf{P}}_\mathbf{n}$ is estimated accurately enough. However, when there are TAD errors, we will show that the performance gap reduces between the two methods. The BDLCMV also handles the possible power leakages of the interferers, and can suppress the interferers that are not included in the constraints.

If each node has only one microphone, then BDLCMV becomes diagonal. In this case, it can be viewed as a weighted version of the LCDS beamformer, and without nulling constraints, can be viewed as a weighted DS beamformer.

### C. Distributed Implementation of the Proposed Method

Given a block-diagonal matrix $\bar{\mathbf{P}}$, which can be $\bar{\mathbf{P}}_\mathbf{u}$, $\bar{\mathbf{P}}_\mathbf{n}$ or $\bar{\mathbf{P}}_\mathbf{y}$, and a known constraint matrix $\mathbf{\Lambda}$, we now demonstrate how we can form a distributed version of the proposed methods for use in general cyclic networks by using a similar technique to that presented in [16]. Importantly, the imposed block diagonal structure of the estimated CPSDM results in a naturally separable objective function, leading to a substantial reduction in communication costs compared to those in [16]. To demonstrate this, denote by $\mathbf{w}_\kappa$, $\mathbf{\Lambda}_\kappa$ and $\bar{\mathbf{P}}_\kappa$ the elements of $\mathbf{w}$, the rows of $\mathbf{\Lambda}$ and the block diagonal component of $\bar{\mathbf{P}}$ associated with node $\kappa$, respectively. Eqs. (11), (12) and (13) can therefore be rewritten as

$$\hat{\mathbf{w}} = \arg\min_{\mathbf{w}} \frac{1}{2}\sum_{\kappa=1}^{N} \mathbf{w}_\kappa^H \bar{\mathbf{P}}_\kappa \mathbf{w}_\kappa \text{ s.t. } \sum_{\kappa=1}^{N} \mathbf{w}_\kappa^H \mathbf{\Lambda}_\kappa = \mathbf{f}^H. \quad (14)$$

The real-valued Lagrangian of this problem is given by

$$\mathcal{L}(\mathbf{w}, \boldsymbol{\mu}) = \sum_{\kappa=1}^{N} \left( \frac{\mathbf{w}_\kappa^H \bar{\mathbf{P}}_\kappa \mathbf{w}_\kappa}{2} - \Re\left(\boldsymbol{\mu}^H \left(\mathbf{\Lambda}_\kappa^H \mathbf{w}_\kappa - \frac{\mathbf{f}}{N}\right)\right)\right),$$

where we have partitioned the constraint vector $\mathbf{f}$ into $N$ equal parts, $\mathbf{f}/N$, one for each node $i \in V$. Taking complex partial derivatives [42], it follows that

$$\hat{\mathbf{w}}_\kappa = \bar{\mathbf{P}}_\kappa^{-1} \mathbf{\Lambda}_\kappa \boldsymbol{\mu}, \quad (15)$$

such that the corresponding dual function is thus given by

$$q(\boldsymbol{\mu}) = -\sum_{\kappa=1}^{N} \frac{\boldsymbol{\mu}^H \mathbf{\Lambda}_\kappa^H \bar{\mathbf{P}}_\kappa^{-1} \mathbf{\Lambda}_\kappa \boldsymbol{\mu}}{2} + \Re\left(\boldsymbol{\mu}^H \mathbf{f}\right).$$

The resulting dual optimization problem is given by

$$\hat{\boldsymbol{\mu}} = \arg\min_{\boldsymbol{\mu}} \sum_{\kappa=1}^{N} \left( \frac{\boldsymbol{\mu}^H \mathbf{\Lambda}_\kappa^H \bar{\mathbf{P}}_\kappa^{-1} \mathbf{\Lambda}_\kappa \boldsymbol{\mu}}{2} - \Re\left(\boldsymbol{\mu}^H \frac{\mathbf{f}}{N}\right)\right). \quad (16)$$

### D. Acyclic Implementation via Message Passing

We begin by demonstrating how, when the underlying network is acyclic (tree structured), the problem in Eq. (16) can be solved in a distributed manner. Similar to the approach introduced in [18], there is no need for this acyclic network to be constant between frames, allowing it to adapt to the time-varying connectivity of dynamic networks. This contrasts [14], [15] where the network topology must remain constant.

In the following, we consider two different approaches to compute the optimal $\boldsymbol{\mu}$ in tree structured networks. In the first approach, we exploit the fact that Eq. (16) can be directly solved by aggregating the sum of the local matrices $\frac{1}{2}\mathbf{\Lambda}_\kappa^H \bar{\mathbf{P}}_\kappa^{-1} \mathbf{\Lambda}_\kappa$ to a common location. In the case of acyclic networks, this aggregation can be performed efficiently with the common location forming the root node of the network. This root node can simply be a point in the network where we choose to extract the beamformer output signal.

To sketch the process of this data aggregation, we partition the set of neighbors of each node $\kappa$ into two groups. The first group, denoted by $\mathcal{C}_\kappa$, represents the set of children of node $\kappa$. The second set, which is a unique node identifier, is the parent of node $\kappa$ denoted by $\mathcal{P}_\kappa$. In particular, $\mathcal{P}_\kappa \cup \mathcal{C}_\kappa = \mathcal{N}(\kappa) \forall \kappa \in V$, where $\mathcal{N}(\kappa) = \{\iota \mid (\kappa, \iota) \in E\}$. Note that for the root node $\mathcal{P}_\kappa = \emptyset$. These sets can be determined per frame by selecting a root node and forming a spanning tree via a breadth-first or depth-first search.

Once these sets are known, the process begins at the leaf nodes of the networks (those nodes for which $\mathcal{C}_\kappa = \emptyset$) and consists of the transmission of a message from these nodes ($\kappa$) to their parents ($\mathcal{P}_\kappa$). The aggregation messages are matrices and take the form

$$\mathbf{M}_{\kappa \to \mathcal{P}_\kappa} = \frac{\mathbf{\Lambda}_\kappa^H \bar{\mathbf{P}}_\kappa^{-1} \mathbf{\Lambda}_\kappa}{2}.$$

Of the set of remaining nodes, those nodes which have received a message from all but one of their neighbors can repeat this process (the remaining neighbor is their parent node). Their messages take a more general form given by

$$\mathbf{M}_{i \to \mathcal{P}_i} = \frac{\mathbf{\Lambda}_i^H \bar{\mathbf{P}}_i^{-1} \mathbf{\Lambda}_i}{2} + \sum_{k \in \mathcal{C}_i} \mathbf{M}_{k \to i},$$

whereby local information at each node is first combined with that from their children. This process is repeated until the root node has received messages from all its children at which point the aggregation operation is complete.

Due to their positive semidefinite structure, the transmission of each message per node comprises $\frac{1}{2}((r+1)^2 + r + 1)$ unique variables resulting in a total of $\frac{1}{2}(r^2 + 3r + 2)(N-1)$ transmitted variables for each frequency bin per frame. The optimal dual variables can then be diffused back into the network to allow the optimal beamformer weight vector to be computed at each node in parallel. This additional diffusion stage results in a further $(r+1)(N-K)$ transmitted variables where $K$ denotes the number of leaf nodes. The beamformer output can then be computed by simply aggregating the sum $\sum_{i \in V} \mathbf{w}_i^H \mathbf{y}_i$ through the network, incurring a total cost of $(N-1)$ transmissions per frequency bin. Finally, if the estimate of $\bar{\mathbf{P}}$ does not change between frames, i.e., $\Delta \bar{\mathbf{P}} = \mathbf{0}$, the estimated weight vector need not be recomputed. An example of this occurs in noisy frames for the proposed BDLCMV method, reducing the cost of this algorithm in such frames to that of simply computing the beamformer output.

### E. Cyclic Weight Vector Computation via PDMM

For more general network structures, Eq. (16) can be transformed to a fully distributable form. To do so, we introduce local versions of $\boldsymbol{\mu}$ at each node, denoted by $\boldsymbol{\mu}_\kappa$, and impose that $\boldsymbol{\mu}_\kappa = \boldsymbol{\mu}_\iota \, \forall (\kappa, \iota) \in E$. The resulting problem is given by

$$\hat{\boldsymbol{\mu}} = \arg \min_{\boldsymbol{\mu}} \sum_{\kappa=1}^{N} \left( \frac{\boldsymbol{\mu}_\kappa^H \mathbf{\Lambda}_\kappa^H \bar{\mathbf{P}}_\kappa^{-1} \mathbf{\Lambda}_\kappa \boldsymbol{\mu}_\kappa}{2} - \Re\left(\boldsymbol{\mu}_\kappa^H \frac{\mathbf{f}}{N}\right) \right)$$
$$\text{s.t. } \boldsymbol{\mu}_\kappa = \boldsymbol{\mu}_\iota \quad \forall (\kappa, \iota) \in E. \quad (17)$$

Note that at optimality, this problem is entirely equivalent to the problem in Eq. (16), assuming the network is connected. Due to its separable quadratic structure, Eq. (17) can be solved via a wide range of existing distributed solvers [43]–[45]. In this work, we consider solving Eq. (17) using the primal dual method of multipliers (PDMM) proposed in [45].

To define the PDMM updating scheme, we begin by again considering the equivalent graph representation of the network, parameterised by node set $V$ and edge set $E$. For each node $\kappa$ and edge $(\kappa, \iota) \in E$, define the vectors $\boldsymbol{\mu}_\kappa^{(0)} = \boldsymbol{\gamma}_{\kappa, \iota}^{(0)} = \mathbf{0} \in \mathbb{C}^{r+1}, \, \forall \kappa = 1, \ldots, N, \, (\kappa, \iota) \in E$ respectively. As per the PDMM algorithm in [45], the optimizers of Eq. (17) can then be computed by iteratively updating the dual variables ($\boldsymbol{\mu}_\kappa$) and directed edge variables ($\boldsymbol{\gamma}_{\kappa|\iota}$) as

$$\boldsymbol{\mu}_\kappa^{(t+1)} = \left( \frac{\mathbf{\Lambda}_\kappa^H \bar{\mathbf{P}}_\kappa^{-1} \mathbf{\Lambda}_\kappa}{2} + \rho |\mathcal{N}(\kappa)| \mathbf{I} \right)^{-1}$$
$$\left( \frac{\mathbf{f}}{N} - \sum_{\iota \in \mathcal{N}(\kappa)} \left( \frac{\kappa - \iota}{|\kappa - \iota|} \boldsymbol{\gamma}_{\kappa|\iota}^{(t)} - \rho \boldsymbol{\mu}_\iota^{(t)} \right) \right)$$
$$\boldsymbol{\gamma}_{\kappa|\iota}^{(t+1)} = \boldsymbol{\gamma}_{\iota|\kappa}^{(t)} - \rho \frac{\kappa - \iota}{|\kappa - \iota|} \left( \boldsymbol{\mu}_\kappa^{(t+1)} - \boldsymbol{\mu}_\iota^{(t)} \right), \quad (18)$$

where each $\rho \in (0, +\infty)$ is the step size for the iterative algorithm and $t$ denotes the iteration index. The notation $\kappa|\iota$ is used to define the edge variable computed at node $\kappa$ related to the edge $(\kappa, \iota) \in E$.

The edge based update requires the transmission of information between neighbouring nodes, as can be noted in the dependence of $\boldsymbol{\gamma}_{\kappa|\iota}^{(t+1)}$ on $\boldsymbol{\gamma}_{\iota|\kappa}^{(t)}$ and $\boldsymbol{\mu}_\iota^{(t)}$. As highlighted in [45] however, this only requires the transmission of the $\boldsymbol{\mu}_\kappa$ variables and, thus, can be performed via a broadcast transmission protocol at each node. These updates can then be iterated until a desired level of precision is achieved after which $\hat{\mathbf{w}}_j$ can be calculated locally at each node via Eq. (15).

Each iteration of the proposed algorithm requires the transmission of $r + 1$ variables per node. In an existing optimal cyclic beamformer [16] this cost was $r + 1 + |L_y|$, where $|L_y|$ is the number of frames used to form a maximum likelihood estimated version of the CPSDM. The proposed method therefore requires $|L_y|$ less transmissions per iteration, resulting in a substantial saving in transmission costs.

### F. Beamformer Output Computation

Once the weight vector is known, the beamformer output can then be computed via various distributed averaging techniques (see [46] for an overview). In the case of this work we again consider the use of PDMM for this task. Consider the standard distributed averaging problem given by

$$\min_{\mathbf{x}} \quad \frac{1}{2} \sum_{\kappa=1}^{N} \|\mathbf{x}_\kappa - \mathbf{w}_\kappa^H \mathbf{y}_\kappa\|^2 \quad (19)$$
$$\text{s.t.} \quad \mathbf{x}_\kappa = \mathbf{x}_\iota \, \forall (\kappa, \iota) \in E.$$

Again, from [45], the PDMM update equations for this problem are given by

$$\mathbf{x}_\kappa^{(t+1)} = \frac{\left( \mathbf{w}_\kappa^H \mathbf{y}_\kappa - \sum_{\iota \in \mathcal{N}(\kappa)} \left( \frac{\kappa - \iota}{|\kappa - \iota|} \mathbf{z}_{\kappa|\iota} - \rho \mathbf{x}_\iota^{(t)} \right) \right)}{1 + \rho |\mathcal{N}(\kappa)|} \quad (20)$$

$$\mathbf{z}_{\kappa|\iota}^{(t+1)} = \mathbf{z}_{\iota|\kappa}^{(t)} - \rho \frac{\kappa - \iota}{|\kappa - \iota|} \left( \boldsymbol{\mu}_\kappa^{(t+1)} - \boldsymbol{\mu}_\iota^{(t)} \right), \quad (21)$$

where $\mathbf{z}_{\kappa|\iota}$ denotes the directed edge variable owned by node $\kappa$. By iterating these updates, every node in the network can learn the average of the vector $\mathbf{w}^H \mathbf{y}$. Once the average is known, this can be scaled by a factor of $N$ to recover the beamformer output. Alternatively, we can employ the same acyclic beamformer output computation approach as used in Sec. V-D. While this removes the entirely cyclic nature of the algorithm as the tree structured network used can change in each frame, the overhead of using an acyclic network is still substantially reduced in contrast to the work of [14], [15].

*G. Cyclic Beamforming with Finite Numbers of Iterations*

In general distributed applications, deterministic signal processing is desirable. This point is even more pressing in the case of distributed audio processing. Thus, an unbounded requirement on the iteration count of an algorithm is cumbersome. Unfortunately, in practice, the total number of transmissions required to solve the problems in Eq. (17) and (19), via general cyclic solvers such as PDMM, is dependent not only on the choice of the solver but also on the WASN topology. As such, it is not possible to analytically bound this transmission cost for arbitrary networks. However, in the distributed beamforming method presented in [26], which also used PDMM as a solver, it was found that near optimal performance was achieved in only a limited number iterations. In this way it is expected that the number of iterations required to achieve a good level of performance is not unnecessarily large. As such we can impose a hard limit on the number of iterations performed without significantly degrading performance.

An additional observation is that, due to its dependence on a recursively averaged covariance matrix, the weight vector **w** will vary smoothly with time. With regards to the PDMM algorithm, this corresponds to the fact that both the dual and edge variables will also vary somewhat smoothly. As such, one way to improve precision even under the scenario of a finite number of iterations it to use a warm-start procedure. Defining the maximum number of iterations by $t_{\max}$, this warm-start procedure is implemented by setting

$$\boldsymbol{\mu}_\beta^{(0)} = \boldsymbol{\mu}_{\beta-1}^{(t_{\max})} \quad \text{and} \quad \boldsymbol{\gamma}_{\beta,\kappa|\iota}^{(0)} = \boldsymbol{\gamma}_{\beta-1,\kappa|\iota}^{(t_{\max})}, \qquad (22)$$

where the additional subscript denotes the frame index $\beta$. In the case of a constant CPSDM estimate this procedure allows the finite iterations in multiple frames to be used to solve the same problem i.e. a higher precision weight vector can be achieved. In the case of slowly varying weight vectors, this allows the algorithm to track the optimal weight vector while still only incurring a finite iteration cost per frame.

A warm-start procedure cannot be used in the case of the beamformer output computation as it varies rapidly between frames. However, only a finite number of iterations are required per frame to achieve near-optimal performance. Thus, an iteration limit can be imposed to achieve a fully cyclic implementation. The performance of this iteration-limited output computation and the warm-started weight vector computation introduced above are demonstrated in Sec. VI-D.

*H. Comparing the Transmission Costs of Different Beamformer Implementations*

Table I includes the transmission costs of the distributed implementations of the BDLCMV/BDLCMP algorithm proposed in this paper. It is worth noting that these transmission costs do not include the additional overhead associated with those algorithms which exploit a TAD or the costs of forming a spanning tree. However, due to the per frequency bin nature of the algorithm, these costs are assumed to be far lower than those associated with running the algorithm.

TABLE I: Transmission costs of distributed beamformers in dynamic sound fields. $N$ denotes the number of nodes, $K$ denotes the number of leaf nodes, $r$ denotes the number of interferers, and $t_{\max}$ denotes the maximum number of iterations.

| *Beamformer Weight Vector Computation* | |
| --- | --- |
| Algorithm | Transmissions per frame & frequency bin |
| BDLCMV/BDLCMP (Cyclic) | $t_{\max}(r+1)N$ |
| BDLCMV/BDLCMP (Acyclic) | $\frac{1}{2}(r^2+3r+2)(N-1)+(r+1)(N-K)$ |
| BDLCMV (Acyclic $\Delta\bar{\mathbf{P}} = \mathbf{0}$) | 0 |
| DLCMV (Acyclic) [14] | $(2N-1-K)$ |
| DGSC (Acyclic) [15] | $(2N-1-K)+(r+1)(N-K)$ |
| TI-DANSE (Cyclic) [18] | $(2N-1-K)(r+1)$ |
| *Beamformer Output Computation* | |
| Algorithm | Transmissions per frame & frequency bin |
| Cyclic | $t_{\max}N$ |
| Acyclic | $N-1$ |

From Table I, our proposed acyclic implementation appears to require a notable increase in total transmission cost when we allow $\bar{\mathbf{P}}$ to vary. However unlike existing approaches, it does so while ensuring we exactly solve the problem in each frame. In contrast, the alternative methods listed require multiple frames to reach optimality [47]. As such, the proposed acyclic approach offers a competitive advantage as it exactly attains the performance of a centralized implementation in each frame while incurring a fixed transmission cost. In contrast, the iterative nature of DLCMV, DGSC and TI-DANSE means that they require multiple frames to achieve the same precision, essentially scaling their effective transmission costs.

The proposed cyclic implementation of BDLCMV/BDLCMP, like other existing approaches within the literature [14], [15] allows for a tradeoff between per-frame optimality and communication overhead. Importantly, when combined with the warm-start procedure introduced in Eq. (22), this allows for near-optimal performance while reducing the total transmission overhead per frame. In particular, in Sec. VI-D we will demonstrate the effect of combining this warm-start procedure with a single iteration, that is $t_{\max} = 1$. In this case, a negligible decrease in performance is achieved while incurring a transmission cost in line with existing acyclic distributed beamformers.

Finally, by providing two methods of beamformer output computation, we allow designers to implement a fully cyclic beamforming algorithm if they desire. Perhaps more attractive though is a hybrid style approach, similar to that used in [18], which combines cyclic weight vector computation with an acyclic output computation stage. This takes advantage of the transmission savings of both approaches while, as the acyclic topology can vary between frames, removes the need for acyclic network management in contrast to [14], [15].

TABLE II: Summary of compared linearly constrained beamformers which are all special cases of the optimization problem in Eq. (5). Note that $\mathbf{w}^H \mathbf{\Lambda} = \mathbf{f}^H$ is the constraints in Eq. (6).

| Method | $\mathbf{P}$ | Constraints | Target activity detection |
|---|---|---|---|
| MPDR | $\mathbf{P_y}$ | $\mathbf{w}^H \mathbf{a} = 1$ | no |
| MVDR | $\mathbf{P_n}$ | $\mathbf{w}^H \mathbf{a} = 1$ | yes |
| DS | $\mathbf{I}$ | $\mathbf{w}^H \mathbf{a} = 1$ | no |
| LCMP | $\mathbf{P_y}$ | $\mathbf{w}^H \mathbf{\Lambda} = \mathbf{f}^H$ | no |
| LCMV | $\mathbf{P_n}$ | $\mathbf{w}^H \mathbf{\Lambda} = \mathbf{f}^H$ | yes |
| LCDS | $\mathbf{I}$ | $\mathbf{w}^H \mathbf{\Lambda} = \mathbf{f}^H$ | no |
| BDLCMP | $\bar{\mathbf{P}}_\mathbf{y}$ | $\mathbf{w}^H \mathbf{\Lambda} = \mathbf{f}^H$ | no |
| BDLCMV | $\bar{\mathbf{P}}_\mathbf{n}$ | $\mathbf{w}^H \mathbf{\Lambda} = \mathbf{f}^H$ | yes |

## VI. EXPERIMENTAL RESULTS

We compare the performance of the proposed beamformers (except of the BDALCMV, where an estimate of $\bar{\mathbf{P}}_\mathbf{u}$ is difficult to obtain), and six existing centralized beamformers (the MPDR, MVDR, LCMP, LCMV, LCDS and DS) in terms of noise suppression, predicted intelligibility improvement, robustness to RATF estimation errors and TAD errors. Table II summarizes the compared linearly constrained beamformers. Note that the ALCMV and ILCMV are not included in the comparisons since there are no distributed estimation methods of $p_{\text{iso}}$. Note that the MPDR, MVDR, LCMP, LCMV, LCDS and DS are distributable under the distributed LCMV (DLCMV) [14], as well as the distributed DS beamformer proposed in [9]. Specifically, we examine the performance of centralized implementations of the aforementioned beamformers to which their distributed counterparts converge [14].

### A. Experiment Setup

The simulations are conducted in a simulated reverberant environment with reverberation times $T_{60} = 0.2$ s and $T_{60} = 0.5$ s using the image method [48]. A box-shaped room with dimensions $6 \times 4 \times 3$ is selected for the reverberant environment. The configuration of the nodes and acoustic sources are depicted in Fig. 3. We considered an example scenario where a number of people are sitting around a table with a set of mobile phones on the table, each equipped with multiple microphones. In this case, $N = 5$ nodes were placed on a virtual surface (with no physical properties) and four sources were placed around the surface. Each node was equipped with 3 microphones forming a uniform linear array with an inter-microphone distance of 2 cm. This resulted in a total of $M = 15$ microphones. Three of the four sources were interferering talkers (2 female and 1 male) with the remainder being the target source (a male talker). Each signal had a simulated duration of 30 s and was sampled at $f_s = 16$ kHz. The power of each interferer at its original position was set to be approximately equal to the power of the target source at its original position (i.e., a 0 dB SNR). The impulse responses between microphones and sources were computed using the toolbox in [49], with length 200 ms. The closest microphone to the target was selected as the reference microphone (see Fig. 3). The microphone-self noise was white Gaussian noise with 40 dB SNR with respect to the target signal at the reference microphone.

As can be noted in Fig. 3, the distance between any two nodes was quite big (i.e., the distance between the closest microphone-pair, where the two microphones belonged to two different nodes, was at least $0.5091$ m). Thus, the ambient noise was approximately spatially uncorrelated between different nodes. As explained in Section II, the late reverberation, which is the main contribution in the ambient noise component, becomes approximately uncorrelated between two microphones with distance $d$ above a certain threshold $f_c = c/(2d)$. Here, the distance of the closest microphone-pair where the microphones belong to two different nodes is 0.5091 m corresponding to $f_c = 333.9$ Hz (if $c = 340$ m/s). Note that the correlation between any other microphone-pair with microphones in different nodes will have even smaller $f_c$.

On the other hand, the late reverberation for microphones within a node is highly correlated. The distance between two consecutive microphones is $d = 0.02$ m and, resulting in $f_c = 8.5$ kHz, which is greater than $f_s/2 = 8$ kHz.

### B. Processing

STFT frame-based beamforming was performed using an overlap and save (OLS) procedure [50]. We used a rectangular analysis window with length $2L_{\text{fr}} = 50$ ms, where $L_{\text{fr}} = 25$ ms is the length of the current frame. Thus, the early-reverberant RATF vectors of the sources are associated with an impulse response of length 50 ms. The analysis window was applied on the current frame and the previous frame in order to a) mitigate circular convolution problems, and b) to be able to handle large phase shifts in the constraints due to the large microphone array aperture. The FFT length is $\Phi = 1024$.

In order to achieve a smoother processing than standard OLS, the analysis window was shifted by $L_{\text{fr}}/2$ samples[2]. A Hann window (synthesis window) was then applied, with length $L_{\text{fr}}$, on the last $L_{\text{fr}}$ processed samples. Finally, the last $L_{\text{fr}}/2$ processed samples were saved in order to add them to the corresponding samples of the next windowed segment.

The CPSDMs, for the $k$-th frequency bin and $\beta$-th analysis segment, were estimated via recursive averaging as described in Section III-B. Note that the block-diagonal CPSDMs were recursively averaged locally at each node. The noise CPSDM and the block-diagonal noise CPSDM were estimated using an ideal TAD and a non-ideal state-of-the-art voice activity detector proposed in [51]. For simplicity, the TAD decision is based only on the reference microphone signal.

The RATF vectors were estimated once using additional 2 s recordings per source. Specifically, each talker spoke alone for 2 s, while all the others were silent. The CPSDM matrices of each talker were computed as described in Section III-B and the dominant relative eigenvector from each CPSDM was selected as an estimate of the RATF vector for each source[3].

---
[2]The standard OLS procedure usually shifts the analysis window by $L_{\text{fr}}$.
[3]If there is a noise component which is always active, such as an air-condition, a more accurate method of estimating the RATF of the talkers is by using the GEVD approach [32].

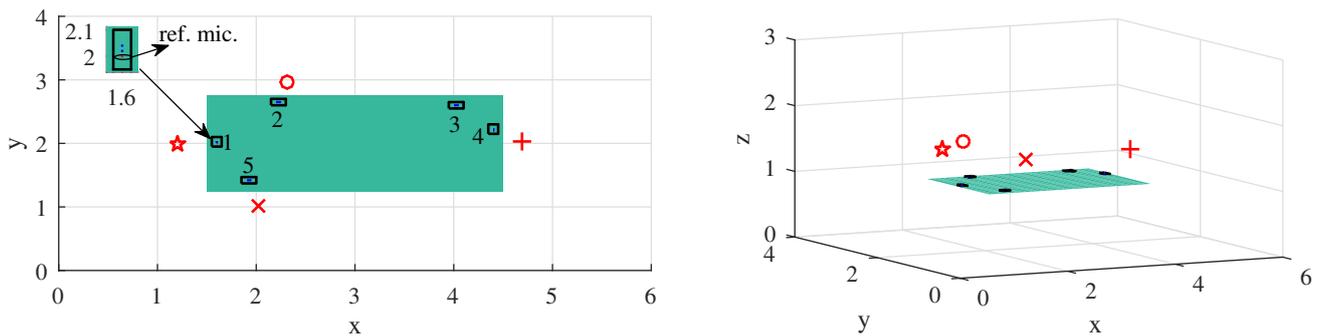

Fig. 3: Experimental setup from two different angles: three interferers (two female talkers with markers '+' and 'x' and one male talker with marker 'o'), one target (a male talker with marker ⋆), and five nodes, with three microphones each, sitting on the virtual surface. The height of the virtual surface is 1 m.

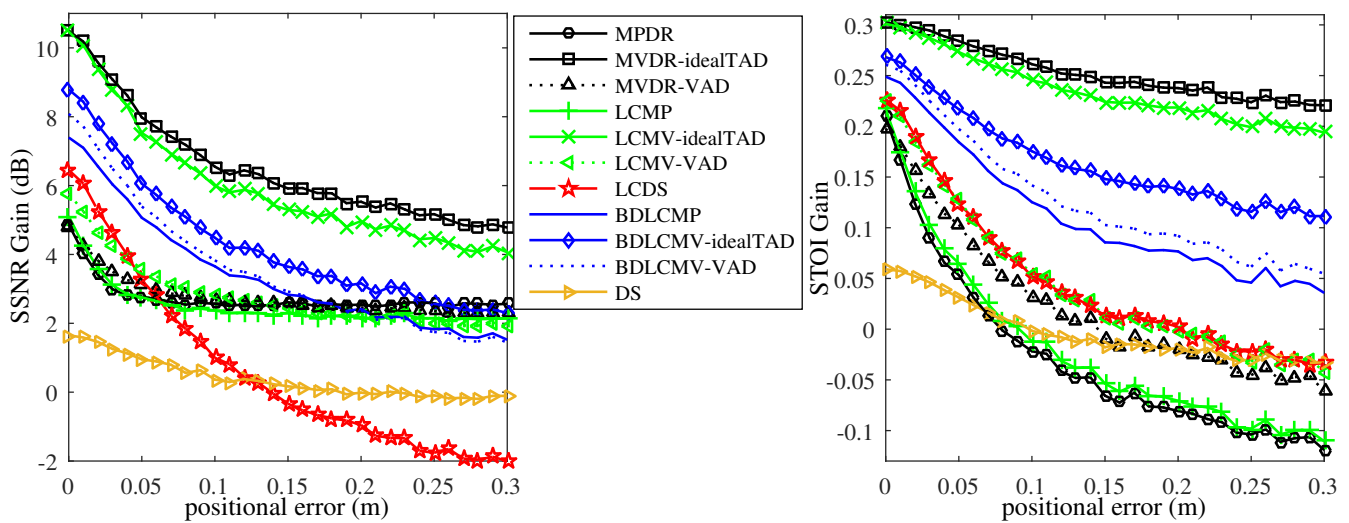

Fig. 4: Reverberation time $T_{60} = 0.2$ s: Comparison of the beamformers in Table II as a function of positional error between training and testing positions. The methods that depend on a TAD are computed using an ideal TAD and the state-of-the-art voice activity detector (VAD) proposed in [51].

These initial positions of the talkers, in which the RATF vectors were estimated, will be referred to as training positions and were nearby to the testing positions depicted in Fig. 3. Therefore, the RATF estimation errors of all sources can be modeled as a function of positional error between the training positions and the testing positions.

### C. Robustness to RATF estimation errors

Figs. 4 and 5 show the performance of the aforementioned beamformers in terms of segmental-signal-to-noise-ratio (SSNR) gain and the short-time objective intelligibility measure (STOI) [52] gain as a function of positional error for $T_{60} = 0.2$ s and $T_{60} = 0.5$ s, respectively. Note that the noise that is computed in the SSNR consists of the interferers, background, and target distortion noise. The erroneous training locations were uniformly distributed over a sphere centered around the true source locations having a radius ranging from $0 - 0.30$ m in 0.01 m steps. For every value of positional error, the average performance of 20 different setups was measured. Each setup used the same source signals at the same testing locations as shown in Fig. 3. However, a different set of initial training positions, computed as mentioned previously, were used in each setup. Likewise, different realizations of the microphone-self noise were also used in each setup.

It is clear that the proposed beamformers are more robust for the combination of large positional and TAD errors. Specifically, the BDLCMV and the BDLCMP provide significantly better predicted intelligibility improvement compared to all the other methods using a non-ideal TAD or not using a TAD. The BDLCMV with the non-ideal TAD is slightly better than the BDLCMP. Thus, in this particular scenario a TAD is not necessary for the proposed method, since it will create errors and the performance advantage will be small. Note that for $T_{60} = 0.5$ s and for large positional errors, the proposed methods achieve worse noise reduction, but better intelligibility improvement, than the other methods. As explained in Section V, this is because the proposed beamformers distort the target signal much less than the other beamformers.

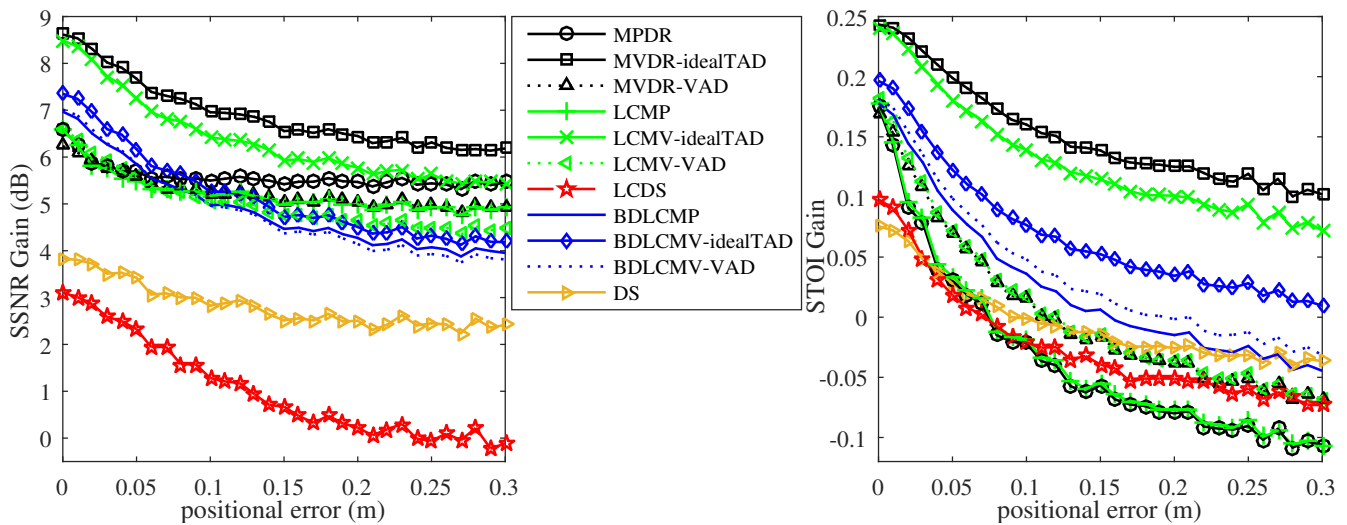

Fig. 5: Reverberation time $T_{60} = 0.5$ s: Comparison of the beamformers in Table II as a function of positional error between training and testing positions. The methods that depend on a TAD are computed using an ideal TAD and the state-of-the-art voice activity detector (VAD) proposed in [51].

The LCMV using the non-ideal TAD is much more robust than the LCMP and gives much higher predicted intelligibility improvement. It is worth noting that for $T_{60} = 0.2$ s the fixed LCDS has almost the same predicted intelligibility improvement as the LCMV. This makes the usage of the LCMV beamformer, in this particular acoustic scenario, obsolete in the distributed context since LCDS has significantly lower communication costs. On the other hand, for $T_{60} = 0.5$ s the performance of LCDS deteriorates significantly and becomes also worse compared to the DS beamformer. Moreover, the MVDR using a non-ideal TAD has almost the same predicted intelligibility improvement with the LCMV using the non-ideal TAD for $T_{60} = 0.5$ s.

In conclusion, for those simulations using a non-ideal TAD, the proposed methods are the most robust out of those considered. Moreover, the proposed method incurs lower communication costs, as explained in Section V, making it a strong candidate for distributed beamforming.

### D. Limiting Iterations per Frame for PDMM Based BDLCMP/BDLCMV

We now compare the impact of a finite iteration cap on the optimality of both the computed beamformer weight vector and beamformer output signal. For these simulations, the same setup, as introduced in Sec. VI-A, was used. The case of BDLCMP with no RATF estimation errors was considered where by the centralized beamformers used previously were substituted with their cyclic counterparts introduced in Sec. V-E. For these simulations, three standard network configurations (a chain, a ring and a star network) were considered to highlight the impact network topology can play on convergence. Examples of these three network topologies are included below in Figures 6a, 6b, 6c respectively. A step size of $\rho = \frac{1}{2}$ was heuristically selected for all simulations. With a more refined selection of this parameter, we expect that faster convergence could be achieved.

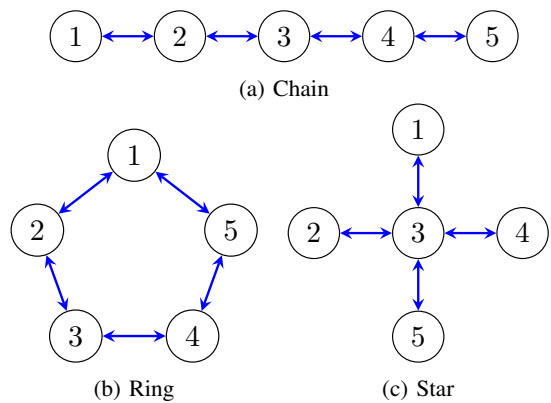

Fig. 6: Chain, Ring and Star topologies for the considered five node network.

Fig. 7 shows a comparison of convergence rates of both cold and warm-started beamformer weight vector computation for the three networks considered. As expected, while all three methods require many iterations ($> 30$) to achieve reasonable weight vector estimation, when combined with a warm-start procedure, even a single iteration per frame achieves near optimal gains in both STOI and SSNR. Thus, for slowly varying CPSDM estimates, the cyclic BDLCMP/BDLCMV approach offers an opportunity to dramatically reduce transmission costs while maintaining near optimal performance. Furthermore, the effectiveness of this warm-start does not seem to vary significantly with network topology.

For beamformer output computation, as demonstrated in

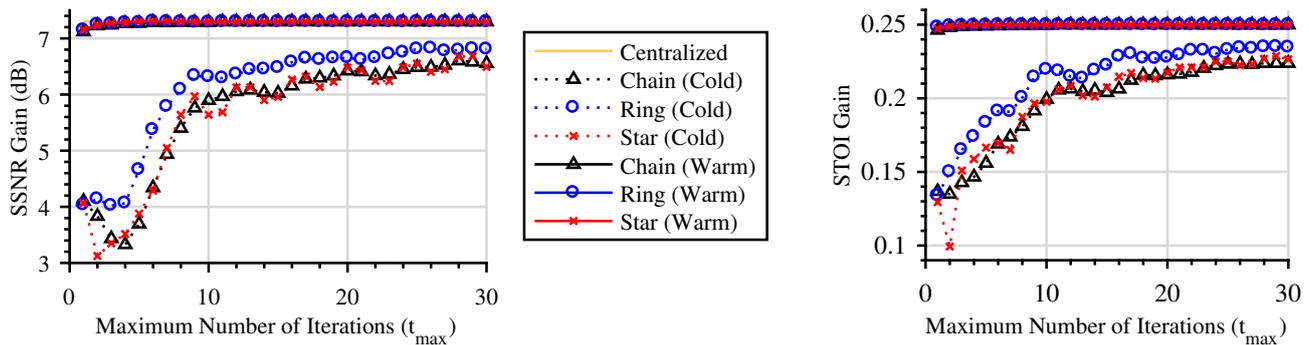

Fig. 7: Comparing the effect of a finite iteration limit on PDMM beamformer weight vector computation. Cold-start (cold) and warm-start (warm) scenarios are considered with the beamformer output being computed exactly via acyclic data aggregation.

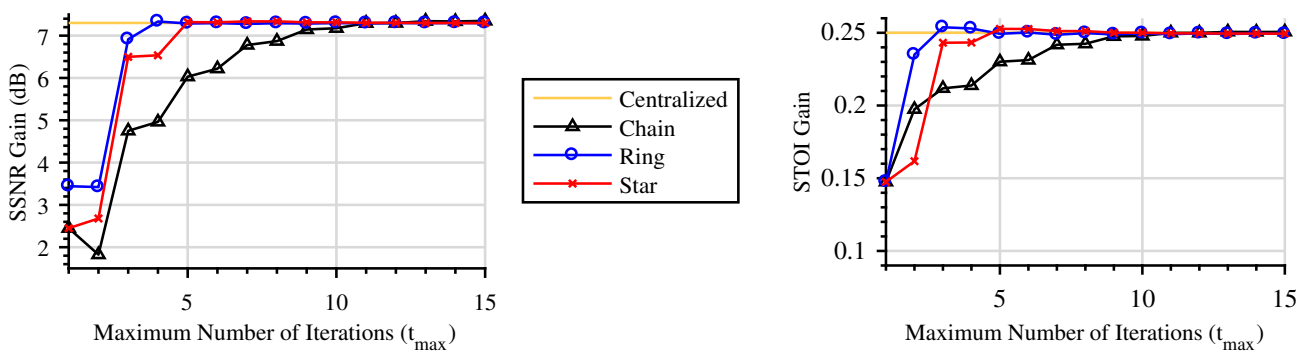

Fig. 8: Comparing the effect of a finite iteration limit on PDMM beamformer output computation. For each of the networks considered the beamformer weight vector is computed exactly via acyclic data aggregation.

Fig. 8, the story is similar. While the dynamic nature of the beamformer output does not facilitate a warm-start procedure, the simplicity of the problem means that within 10 iterations or so a near optimal beamformer output is computed.

Unlike the beamformer weight vector computation, here we can more clearly observe the effect of network topology on convergence. In particular, the chain network, which has a larger diameter than either the ring or the star network, requires roughly twice the number of iterations to approach optimal convergence. This point is consistent with the fact that an even length chain network has twice the diameter of a ring network of the same size. However, this may be able to be remedied with more careful step size selection.

## VII. CONCLUSION

In this paper, we proposed a new distributed linearly constrained beamformer, which provides increased robustness to TAD and RATF estimation errors compared to traditional LCMV-based beamformers. Moreover, the proposed approach is immediately distributable due to its use of a block-diagonal CPSDM. Unlike most competing distributed beamformers, the proposed method can be applied in arbitrary network topologies, while at the same time having much lower communication costs in comparison to competing cyclic approaches and comparable costs to acyclic ones. Furthermore, the general nature of the distributed algorithm facilitates a trade off between transmission costs and per-frame optimality allowing it to be tailored to the needs of a particular application.

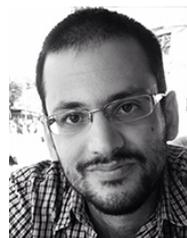
**Andreas I. Koutrouvelis** received the B.Sc. degree in computer science from the University of Crete, Greece, in 2011 and the M.Sc. degree in Electrical Engineering from Delft University of Technology (TU-Delft), the Netherlands, in 2014. From February 2012 to July 2012, he was a research intern at Philips Research, Eindhoven, the Netherlands and from October 2014 to December 2014 he was researcher in the Circuits and Systems Group (CAS) in TU-Delft. Since, January 2015 he is pursuing the Ph.D. degree in TU-Delft (CAS). His research interests include speech analysis and multi-channel speech enhancement.

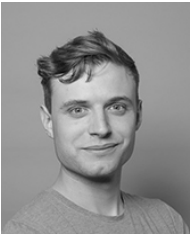
**Thomas W. Sherson** was born on March 30th, 1992 in the town of Peterfield, in Hampshire England. He received his Bachelor of Engineering with First Class Honours, majoring in Electrical and Computer Systems Engineering, from Victoria University of Wellington in New Zealand, in 2015. He was also awarded the Victoria University Medal of Academic Excellence in the same year. Following his graduation, he joined the Department of Microelectronics at Delft University of Technology to continue his studies towards a Doctor of Philosophy (PhD) in the field of Electrical Engineering. His general interests include the likes of signal processing in wireless sensor networks, distributed/decentralised optimisation, monotone operator theory and audio signal processing. Additionally he is an avid outdoorsman with a passion for nature and a love for music.

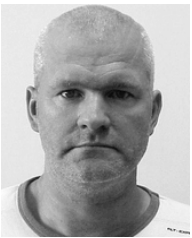
**Richard Heusdens** received the M.Sc. and Ph.D. degrees from Delft University of Technology, Delft, The Netherlands, in 1992 and 1997, respectively. Since 2002, he has been an Associate Professor in the Faculty of Electrical Engineering, Mathematics and Computer Science, Delft University of Technology. In the spring of 1992, he joined the digital signal processing group at the Philips Research Laboratories, Eindhoven, The Netherlands. He has worked on various topics in the field of signal processing, such as image/video compression and VLSI architectures for image processing algorithms. In 1997, he joined the Circuits and Systems Group of Delft University of Technology, where he was a Postdoctoral Researcher. In 2000, he moved to the Information and Communication Theory (ICT) Group, where he became an Assistant Professor responsible for the audio/speech signal processing activities within the ICT group. He held visiting positions at KTH (Royal Institute of Technology, Sweden) in 2002 and 2008 and is a part-time professor at Aalborg University. He is involved in research projects that cover subjects such as audio and acoustic signal processing, speech enhancement, and distributed signal processing for sensor networks.

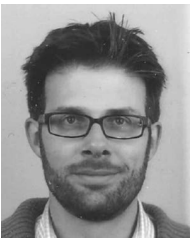
**Richard C. Hendriks** obtained his M.Sc. and Ph.D. degrees (both cum laude) in electrical engineering from Delft University of Technology, Delft, The Netherlands, in 2003 and 2008, respectively. From 2003 till 2007, he was a Ph.D. Researcher at Delft University of Technology, Delft, The Netherlands. From 2007 till 2010, he was a Postdoctoral Researcher at Delft University of Technology. Since 2010, he has been an Assistant Professor in the Signal and Information Processing Lab of the faculty of Electrical Engineering, Mathematics and Computer Science at Delft University of Technology. In the autumn of 2005, he was a Visiting Researcher at the Institute of Communication Acoustics, Ruhr-University Bochum, Bochum, Germany. From March 2008 till March 2009, he was a Visiting Researcher at Oticon A/S, Copenhagen, Denmark. His main research interests are digital speech and audio processing, including single-channel and multi-channel acoustical noise reduction, speech enhancement, and intelligibility improvement.